# Quantum sensing in a physiological-like cell niche using fluorescent nanodiamonds embedded in electrospun polymer nanofibers

J. C. Price[1], S. J. Levett[1], V. Radu[1], D. A. Simpson[2], A. Mogas Barcons[3], C. F. Adams[3], M. L. Mather[1*]

[1]Optics and Photonics Research Group, Faculty of Engineering, University Park, Nottingham, NG7 2RD, UK

[2]School of Physics, University of Melbourne, Parkville, 3010, Australia

[3]School of Life Sciences, Huxley Building, Keele University, Keele, ST5 5BG, UK

*Abstract*

*Fluorescent nanodiamonds (fNDs) containing Nitrogen Vacancy (NV) centres are promising candidates for quantum sensing in biological environments. However, to date, there has been little progress made to combine the sensing capabilities of fNDs with biomimetic substrates used in the laboratory to support physiologically representative cell behaviour. This work describes the fabrication and implementation of electrospun Poly Lactic-co-Glycolic Acid (PLGA) nanofibers embedded with fNDs for optical quantum sensing in an environment, which recapitulates the nanoscale architecture and topography of the cell niche. A range of solutions for electrospinning was prepared by mixing fNDs in different combinations of PLGA and it was shown that fND distribution was highly dependent on PLGA and solvent concentrations. The formulation that produced uniformly dispersed fNDs was identified and subsequently electrospun into nanofibers. The resulting fND nanofibers were characterised using fluorescent microscopy and Scanning Electron Microscopy (SEM). Quantum measurements were also performed via optically detected magnetic resonance (ODMR) and longitudinal spin relaxometry. Time varying magnetic fields external to the fND nanofibers were detected using continuous wave ODMR to demonstrate the sensing capability of the embedded fNDs. The potential utility of fND embedded nanofibers for use as biosensors in physiological environments was demonstrated by their ability to support highly viable populations of differentiated neural stem cells – a major therapeutic population able to produce electrically active neuronal circuits. The successful acquisition of ODMR spectra from the fNDs in the presence of live cells was also*

*demonstrated on cultures of differentiating neural stem cells. This work advances the current state of the art in quantum sensing by providing a versatile sensing platform that can be tailored to produce physiological-like cell niches to replicate biologically relevant growth environments.*

The Nitrogen-Vacancy (NV) color center in diamond has emerged as a high performance quantum sensor with application areas including photonics, quantum information science and the life sciences [1,2,3]. The NV color centre is a naturally occurring paramagnetic impurity comprising of a substitutional nitrogen atom adjacent to a vacant lattice site, which in the negative charge state (NV⁻) forms a spin triplet, $m_s = -1, 0$, and $+1$. Upon excitation with green light, the NV⁻ center produces a broadband fluorescent emission extending into the near infrared region with a zero phonon line at 637 nm. The NV⁻ center has attracted attention as a potential fluorescent probe for use in biological imaging due to its high quantum yield, robust luminescence, which does not bleach or blink under normal conditions, and the inherent low cytotoxicity of diamond[3]. The ability to optically manipulate and monitor the electron spin state in the NV⁻ center is core to its quantum sensing capabilities. Indeed, optical excitation of the NV⁻ center causes pumping of the spin into the $m_s = 0$ ground state, as a result of spin dependent radiative and non-radiative de-excitation pathway[4]. Due to the axial symmetry of the NV⁻ center the $m_s = \pm 1$ spin levels are degenerate in the absence of a magnetic field and in the ground state are separated from the $m_s = 0$ spin level by an energy of 2.87 GHz[4]. The photoluminescent (PL) intensity from the NV⁻ center is dependent on the electron spin state, which is higher when in the $m_s = 0$ state as compared to the $m_s = \pm 1$ spin levels. Measurement of the PL intensity as a function of microwave (MW) frequency enables acquisition of an optically detected magnetic resonance (ODMR) spectrum. Here, the PL intensity reduces when the MW frequency is resonant with the spin transition from $m_s = 0$ to $m_s = \pm 1$. The structure of the ODMR spectrum is extremely sensitive to external perturbations including magnetic and electric fields, temperature, pressure and lattice strain[3]. This is at the heart of the quantum sensing capabilities of the NV defect, which manifests itself as splitting of the ODMR resonance in the presence of external magnetic fields and shifts in the electronic energy states as a result of electric fields, temperature or strain in the diamond lattice. Quantum sensing is also possible through measurement of the transition rate between the $m_s = 0$ and $m_s = \pm 1$ sublevels, given by the spin-lattice (or longitudinal) relaxation time, $T_1$.

There has been a growing interest in exploiting the spin dependent optical properties of the NV center to probe biological systems, most notably in recent years for opto-magnetic imaging in bacterial cultures[5], magnetic resonance imaging of spin labelled cell membrane proteins[6], optical detection of action potentials in single neurons[7] and intracellular thermometry[8,9]. Despite continuing advances in proof of principle studies using NV quantum sensors in biological systems, there is a paucity of studies addressing the significant challenge of implementing these sensors in appropriate growth environments that recapitulate the *in vivo* cell niche. Currently, many sensing protocols utilise NV centers implanted near the surface of single crystal diamond chips as growth substrates to support monolayer cell culture. Whilst diamond has a low cytotoxicity profile[10], diamond chips lack the intrinsic mechanical, structural and chemical signatures to support functional cell behaviour compared with more native cell niches. Studies using nanodiamonds containing NV centres have targeted individual cells, but again predominantly for the study of cells in monolayer culture. Both of these approaches do not enable replication of important bioactive cues that occur in the three dimensional *in vivo* environment. Indeed, it is well established in the life sciences that a key component of the *in vivo* environment affecting cell function is the extracellular matrix (ECM), as demonstrated by studies reporting the frequency of postsynaptic potentials in neurons to be significantly reduced in a matrix deficient environment[11], reported changes in cell proliferation and gene expression[12], and changes in electrophysiological activity in neuronal[13] and cardiac cells[14] as a result of changes in ECM stiffness.

Significant work has been carried out over the past two decades to establish surrogate ECM environments to support *in vitro* cell culture, this is particularly so in the area of stem cell culture to promote cell function and differentiation and in Tissue Engineering to provide structural cues to guide regenerating tissue organization. Electrospun polymer nanofibers (ENF) are synthetic scaffolds that can replicate essential properties of the ECM to support functional cell behaviour *in vitro*. The fabrication process enables controlled production of nanoscale dimensions and topography similar to the fibrillar structure of the ECM with tuneable fiber thickness, organisation and density. The fibers themselves are typically composed of biodegradable polymers that are amenable to functional modifications to enhance cell survival, differentiation and organisation[15]. ENFs can support myocardial regeneration[16], functional neuronal development[17] and repair of neural injury[18]. Further, due to the low cost and ease of fabrication of ENFs, a large variety of novel biomedical applications have been developed in recent years, such as the controlled release of drugs from nanoparticle embedded ENFs[19,20] and electrical stimulation using conductive ENFs for accelerated maturation of *in vitro* cultured neurons[21].

This work describes the fabrication and implementation of ENFs embedded with fNDs for optical quantum sensing to address the need for biological systems that incorporate high precision biosensors to monitor cell function, together with environments that mimic *in vivo* cellular niches. Specifically, protocols and methodologies are established to embed fNDs into ENFs composed of poly(lactic-co-glycolic acid) (PLGA) nanofibers. The resulting nanofibers were characterised using fluorescent microscopy and Scanning Electron Microscopy (SEM) and the sensing capabilities of the embedded fNDs assessed via ODMR, longitudinal spin relaxometry and detection of external time varying magnetic fields. The growth of neural stem cells (NSCs) on the nanofibers was compared to control cells cultured on planar glass substrates. Herein, a new approach to quantum sensing in live cells is demonstrated in a physiological-like cell niche using fNDs embedded in ENFs.

*Experimental Procedures*

*Fabrication and characterisation of fND embedded electronspun polymer nanofibers*

Aqueous suspensions of 100 nm diameter fNDs containing a high dose of NV$^-$ centres (>1000/particle) at a concentration of 1 mg/mL were purchased from FND Biotech, Inc. Solutions for electrospinning were prepared by first centrifuging 1 ml of the fND suspension at 21,000 x g for 10 minutes. Next, the supernatant was removed to produce a pellet of fNDs, which was subsequently dried under a nitrogen atmosphere to remove residual water. A range of candidate solutions for electrospinning were prepared by adding a single fND pellet to 0.33 ml solutions of absolute methanol, acetonitrile or ethyl acetate. Each solution was vortexed and added to 0.67 ml solutions of high grade chloroform (1:2 ratio). 80 mg of PLGA (Resomer® RG 505, Poly(D,L-lactide-co-glycolide, Sigma)) was then added to each of the resulting solutions. The solutions were mixed using a magnetic stirrer for 1 hour at 20 °C or overnight at 4 °C, producing a 1 ml viscous solution of milky appearance for electrospinning. Critical requirements for preparation of the final solution are use of PLGA that has not undergone degradation due to long storage times and the use of high grade solvents stored on molecular sieves to avoid the presence of water inhibiting the dissolution of PLGA. Prior to electrospinning, the solution was sonicated for one hour to disperse the fNDs enabling uniform distribution of the fNDs in the spun fibers. A conventional electrospinning set up was used[18], briefly this involved loading the solution to be spun into a glass syringe that was then placed on a syringe pump mounted to a mechanical jack for height adjustment. Two electrodes were constructed, consisting of a metal needle attached to the syringe and a grounded aluminium collection plate. The syringe pump was set to extrude the solution at a rate of 0.08 ml/min. Under a sufficiently large voltage, electrostatic repulsion induced stretching of the droplets leaving the syringe, followed by the formation of a jet of polymer toward the collection plate. As the solvent component of the solution evaporates, continuous fibres of polymer are extruded and collected onto glass coverslips adhered to the aluminium electrode, generating a mat of fND embedded nanofibers. Fluorescent dye, Rhodamine-B dye (Sigma), at a concentration of 1 µg/ml was added to each candidate solution for electrospinning (in the absence of fNDs) to enable assessment, by fluorescent microscopy, of the effect of each solvent mixture on the electrospun fibers. The fND embedded ENFs were also imaged using SEM (Hitachi S45). Here fiber coated coverslips were mounted on aluminium stubs and gold sputtered under vacuum for two minutes prior to imaging. Image analysis was performed using the open source plugin FibrilJ[22], to produce histograms of the diameter of fibers and calculate the average fibre diameter in each image. For cell culture work, the coverslips coated with fND embedded ENFs were sealed to 8 well bottomless chambers (Nunc, Sigma) using silicon adhesive and sterilized in a UV chamber (Biorad) for 15 minutes prior to cell seeding.

*Optically detected magnetic resonance*

Experiments were performed using a wide field epi-fluorescence microscope (Olympus IX83) illuminated with a mercury arc lamp. The emission from the lamp was filtered through a narrow band pass excitation filter (Chroma Technology Corporation) centred at 532 nm ± 5 nm. The light was then focused onto the backfocal plane of a 1.49 NA Olympus 60x TIRF objective lens, with the fluorescence emission collected through a 633 nm long pass emission filter (Chroma Technology Corporation) and projected onto a sCMOS camera (Photometrics 95 Prime B) producing an image with a 220 µm x 220 µm field of view (FOV). MWs were generated using a Keysight N2581B vector signal generator and amplified using an AR research 20S1G4 MW amplifier. MW transmission to the fNDs was achieved through a 0.125 mm diameter copper wire positioned above the ENF coated coverslip mounted on the microscope stage. Instrumentation and acquisition settings were computer controlled using an Olympus U-RTC control box in combination with Olympus Cellsense imaging software.

Continuous wave ODMR was performed by imaging the fluorescent emission from fNDs in the ENFs whilst sweeping the MW frequency between 2.77 GHz and 2.97 GHz with 2 MHz step sizes. Images were recorded for 10 consecutive frequency sweeps and the resulting data analysed to produce an averaged ODMR spectrum, giving a temporal resolution for a full ODMR spectrum in this instance of 2 seconds. Two point single frequency measurements of the PL intensity were also performed by acquiring two consecutive images, one during the application of a 2.87 GHz MW field and the other without MWs. This measurement scheme was used to dynamically monitor the presence of an externally applied magnetic field. The averaged PL intensity of each image was determined and the difference in the PL intensity related to the magnetic field strength through the use of a calibration curve obtained using a Hall probe. The two point single frequency measurements yield a temporal resolution of 20ms, an effective increase in temporal resolution of 100x compared with a obtaining a full ODMR spectrum. In effect the temporal resolution of the system is limited to 2 times the maximum frame rate of the camera. The max frame rate for a full FOV in this setup is set at 100Hz, however this can be increased to frame rates in excess of kHz if the FOV of acquisition is reduced sufficiently.

*Longitudinal spin relaxometry*

Longitudinal spin relaxation times ($T_1$) of NV centers in the fNDs were determined by measuring the PL intensity as a function of time following initialisation of the spin states by a pulse of green light. Images were acquired using a custom confocal microscope built around a commercial microscope base (Nikon Ti-U). Pulsed optical excitation was achieved using a 532 nm laser (Verdi, Coherent Scientific) in combination with an acousto-optic modulator (Crystal Technologies), controlled by a PulseBlaster acquisition card (Spincore). The PL emission was filtered by a 560 nm long pass filter (Semrock) in combination with a 650 nm to 750 nm bandpass filter (Semrock) and recorded using a single photon counting avalanche photodiode (PerkinElmer), with the time-correlated photons captured with a multiple-event time digitizer card (Fast ComTech P7889). The microscope and data acquisition were controlled via custom LabVIEW code.

*Neuronal cell culture materials*

All cell culture reagents and cell culture grade plastics were purchased from Sigma-Aldrich (Dorset, UK) or Fisher Scientific (Loughborough, UK) unless otherwise stated. Basic fibroblast growth factor (FGF-2) was from Peprotech (Rocky Hill, NJ, USA) and epidermal growth factor (EGF) was from R&D systems Ltd (Abingdon, UK). DNase I was from Roche (Welwyn, UK). The LIVE/DEAD Viability/Cytotoxicity Assay Kit was from Fisher Scientific. Primary antibodies were rabbit anti-β-tubulin (Tuj-1) from Biolegend (Cat: 802001, London, UK), rat anti-myelin basic protein (MBP) from Serotec (Cat: MCA409S, Kidlington, UK) and rabbit anti-glial fibrillary acidic protein (GFAP) from DakoCytomation (Cat: Z033429-2, Ely, UK). Secondary antibodies were from Jackson Immunoresearch Laboratories Ltd (Westgrove, PA, USA). Vectashield mounting medium with 4,6-diamidino-2-phenyl-indole (DAPI) was from Vector Laboratories (Peterborough, UK). The care and use of all animals used for cell culture were in accordance with the Animals (Scientific Procedures) Act of 1986 (UK) with approval by the local ethics committee.

*Neural stem cell derivation, culture and experimental conditions*

Neural stem cells (NSCs) were derived from the subventricular zone of P1-3 CD10 mice as previously described (Adams, Pickard and Chari, 2013; Pickard, Adams and Chari, 2017). Briefly, after dissection, the subventricular zone was dissociated into single cells through addition of 10% DNase and mechanical dissociation. Single cells were suspended in neurosphere medium (3:1 mix DMEM:F12, 2 % B27, 25 ng ml$^{-1}$ EGF and FGF-2, 5 ng ml$^{-1}$ heparin, 50 U ml$^{-1}$ penicillin and 50 µg ml$^{-1}$ streptomycin) at 1 x 10$^5$ viable cells/mL and 5 mL added to T25 flasks for propagation as neurospheres at 37 °C, 5 % $CO_2$.

Flasks were fed every 2 to 3 days. For passaging, neurospheres were dissociated into single cells using a mixture of Accutase and DNase (9:1) and re-seeded in T25 flasks as before. For experiments, glass rectangular coverslips with or without the ENFs were placed in 60 mm petri dishes and coated prior to cell seeding by application of poly-ornithine (20 %, 1 h, 37 °C), rinsing in PBS, application of laminin (5 µg mL-1, 1 h, 37°C) and then rinsing three times with PBS. Single cells (up to a maximum of passage three) were then suspended at $3 \times 10^5$ viable cells/mL in monolayer medium (1:1 mix DMEM:F12, 1% N2, 25 ng ml−1 EGF and FGF-2, 5 ng ml−1 heparin, 50 U ml−1 penicillin and 50 µg ml−1 streptomycin). 3 mL of the suspension was added to each petri dish, which contained one experimental substrate. Once the cells had reached approximately 70 % confluence on the glass substrates (2 to 3 days), the monolayer medium was removed entirely and replaced with differentiation medium (neurosphere medium minus growth factors with addition of 1% foetal bovine serum). Cells were allowed to grow in differentiation medium for seven days with feeding (50% medium change) every 2 to 3 days. Here, cells were either used for the live/dead assay or fixed in 4% paraformaldehyde (20 min at room temperature) for immunocytochemistry.

*Live/dead staining and immunocytochemistry*

Cells were washed once with PBS. DMEM (without additional factors), containing 4 µM calcein AM and 6 µM ethidium homodimer, was then added to the cells. The cells were placed back in the incubator for 15 minutes before immediately being imaged *in situ*. For immunocytochemistry, fixed cells were incubated in blocking solution (5% normal donkey serum [NDS], 0.3% triton-X in PBS) for 30 minutes at room temperature. This was then replaced with appropriate primary antibody solutions which were made up in blocking solution: GFAP for astrocytes, Tuj-1 for neurons and MBP for oligodendrocytes for 16 h at 4°C. Primary antibody solution was washed off using three washes of PBS before addition of blocking solution for 30 minutes at room temperature. The appropriate secondary antibody in blocking solution was then added and incubated for 2 hours at room temperature in the dark. After incubation the samples were washed three times in PBS. Coverslips were then mounted to microscope slides using Vectashield mounting medium.

*Cell imaging and analysis*

Immunocytochemistry imaging was performed using a Leica (Leica Microsystems, Wetzlar, Germany) DM2500 LED fluorescence microscope. Live/dead imaging was performed on a Zeiss AxioObserver.Z (Carl Zeiss MicroImaging mbH, Goettingen, Germany). For quantification of the number of viable cells, three areas from each sample were imaged and green (live) and red (dead) cells were counted with percentage viability calculated as the percentage of green cells of the total number of cells. The proportions of cells displaying cell specific staining and percentage of pyknotic nuclei were also assessed in this way using fixed, immunostained samples. Nuclei associated with cell specific staining were counted as positive marker expression and the percentage of positive nuclei was calculated compared to the total number of nuclei in the image. Pyknotic nuclei were identified as condensed and/or fragmented nuclei, indicative of apoptosis. Column charts are expressed as mean ± standard error of the mean. Statistical differences between the two groups were interrogated by an unpaired, two-sided t-test with a threshold of $P<0.05$ set for any significance.

*Calcium staining and live ODMR*

NSCs were isolated and cultured as previously described, and maintained in culture for 21 days. Prior to imaging, imaging buffer was prepared containing 135mM NaCl, 3mM KCl, 10 mM Hepes, 15 mM glucose, 1mM $MgSO_4$, 2mM $CaCl_2$ after which the PH was set to 7.4 by adding drop wise 1M NaOH. Neuronal cell cultures were left in a biological flow hood for 5 minutes to allow cells to equilibrate to room temperature. Cells were then washed twice in imaging buffer and incubated for 30 minutes at room temperature covered from light in 10µl of 100µM Fluo-5F AM (ThermoFisher) in 20%pluronic acid per 1ml of imaging buffer. Cells were then washed twice in imaging buffer and left in fresh buffer for a further 30 minutes to allow the stain to de-esterify. Cells were then transferred to the microscope (Olympus Ix83) for live imaging and ODMR. Calcium flux in cells was visualised in cells using an Olympus 40x LUCPLFLN 0.6NA objective, with excitation from a mercury arc lamp filtered through a 470-490nm band pass filter, and emission collected through a 515-550nm band pass filtered. Images were captured on an sCMOS camera as previously described. ODMR spectra of the underlying embedded fNDs in live cells were acquired as previously described.

*Results and Discussion*

*Characterisation of fND embedded ENFs*

The composition of solutions for electronspinning notably affected fND aggregation. Visual inspection of solutions prior to electrospinning (Figure 1a) shows the mixture containing a 2:1 ratio of chloroform to methanol to be the most effective combination studied for uniform dispersal of the fNDs. Further, fluorescence microscopy of the Rhodamine-B labelled ENFs, without fNDs, demonstrates the formulation containing a 2:1 ratio chloroform to methanol to produce fibers with a low rate of blebbing as compared to the other solutions studied (Figure 1b). Based on this finding, all subsequent experiments were performed using ENFs spun from a solution containing a 2:1 ratio of chloroform to methanol.

Figure 2 contains a macroscale image of the fND embedded ENFs, images obtained using light microscopy highlighting the location of fNDs in the fibers, along with SEM images and data from image analysis of ENFs with and without fNDs. Light microscopy images clearly demonstrate the uniform distribution of fNDs within the ENFs (Figure 2a). Qualitative inspection of the SEM images indicates the addition of fNDs had little impact on the organisation and size of ENFs (Figures 2b and 2c). This is further evidenced through the comparable average fiber diameters, 365 nm (SD±182nm) without fNDs (Figure 2d) and 306 nm (SD±184nm) with fNDs (Figure 2e). It should also be noted that the final ENF coverage, density, thickness and fiber diameter can be tuned by adjusting experimental parameters such as the distance between the syringe needle and the collection plate, the applied voltage across the electrodes and the spinning time. Overall, the above results demonstrate effective fabrication protocols for uniform embedding of fNDs in ENFs.

The effect on PL and quantum sensing of embedding the fNDs in the ENFs was investigated through a comparison of drop cast fNDs with fND embedded ENFs. Figures 3a and 3b display fluorescent microscopy images obtained from fNDs drop cast on a glass coverslip and fND embedded ENFs (FOV 220 µm x 220 µm). Of note is the low background signal from the PLGA fibers, which are known to be autofluorescent. The PL emission spectra were also collected using a UV-Vis spectrophotometer (Flame, Ocean optics) to compare the emission from the drop cast and embedded fNDs. Although the spectral features are comparable, both cases showing the NV$^-$ zero phonon line at 637 nm. From a quantum sensing point of view, results also demonstrate successful acquisition of an ODMR spectrum from fNDs embedded in ENFs (Figure 3d) which is characterised by a minimum in PL intensity centred around 2.87 GHz, with a split indicative of lattice strain, and a full width at half maximum of 20.3 MHz, consistent with nanodiamonds containing an ensemble of NV centres [23].

*Longitudinal spin relaxation measurements*

Results from measurement of the longitudinal spin relaxation times ($T_1$) of between 40 and 50 individual clusters of fNDs of comparable intensity and size are shown in Figure 4. Overall, in agreement with previous reports[24,25], a broad distribution of $T_1$ relaxation times is observed in all fNDs studied which is likely due to variations in the size, shape and dispersivity of individual fNDs as indicated by histograms (Figure 4b). The results also demonstrate an almost two fold reduction in $T_1$ in the case of embedded fNDs ($T_1$=57 μs ±8 μs) as compared to drop cast fNDs ($T_1$=103 μs ± 9 μs). Multiple effects are envisaged to contribute to this reduction, amongst them differences in the local chemical environment and changes in the relaxation rate of the metastable singlet states[25].

The successful implementation of $T_1$ relaxometry using the fND embedded ENFs provides a pathway to multifunctional NV based quantum sensing in an environment which recapitulates the nanoscale architecture and topography of the cell niche. Protocols previously implemented on NV implanted diamond demonstrating quantification of paramagnetic species in solution, analogous to the detection of endogenous free radicals in living cells could be performed using the fND embedded ENFs[26]. Furthermore, recently reported protocols using polymer encapsulation of fNDs to achieve gadolinium based sensing of pH and redox potentials using $T_1$ relaxometry could also be performed[26]. It is noted that for this application, there is rapid drop off in target signal strength as a function of distance from the ND surface, thus optimisation of ENF fabrication process would need to be considered for such applications. In addition to this, while the sensing capabilities of the fNDs embedded in the ENFs was compromised, through reduction in the $T_1$ relaxation time as compared to bare fNDs, this could be mitigated by investigating different polymer compositions and electronspinning parameters affecting fiber density, thickness, and distribution.

*Time resolved detection of external magnetic fields*

The protocol used for determining the strength of an externally applied DC magnetic field is shown schematically in Figure 5a. The method relies on the comparison of PL intensity acquired with and without MWs (Figure 5b), calculation of the contrast between the corresponding intensities and the direct correlation of the ODMR contrast to the magnetic field strength, as measured by a Hall probe. Using this approach there is no requirement for long NV coherence times, alignment of the NV axis with the applied external field or complicated MW pulse sequences. Moreover, the approach is robust and can readily be implemented in cell culture systems. Experimentally, the protocol involved acquisition of two consecutive images of PL intensity, one with MWs applied at the resonant frequency and one in the absence of a MW field. The applied magnetic field was varied by moving a neodymium permanent magnet mounted on a translation stage above the microscope stage. At each position of the magnet the contrast between the images with and without MWs was calculated using the ratio of the averaged PL intensity in the corresponding images. A calibration curve was then generated by correlating this contrast to magnetic field strength as measured by a transverse Hall probe located on the microscope stage (Figure 5b). The magnetic field was varied between 2 µT and 4mT. Figure 5b shows a dynamic range of up to 1mT, with contrast saturating at higher magnetic field strengths. In this experimental approach we determine the minimum detectable field by taking the gradient of the line at the steepest point in the fitted curve in Figure 4B, yielding a gradient of 3.54% contrast/mT. Considering our error bars in Figure 4b, a ratio of the error to the calculated gradient determined a minimum detectable field of 50±9 µT. Time resolved magnetic field sensing was demonstrated (Figures 5c and 5d) by running the acquisition protocol described above over a period of between 10 and 15 seconds and intermittently placing a neodymium permanent magnet in proximity to the fibres. The PL intensity in the presence of resonant MWs changes when the magnet is placed near the fND embedded ENFs (Figure 5c). A corresponding change in contrast is also seen (Figure 5d) demonstrating the capacity of the fND embedded ENFs for time resolved magnetic field sensing with sub-second temporal resolution of 40 ms.

The above results demonstrate a simplified approach to NV based magnetometry using the fND embedded ENFs. Alternative approaches to magnetometry are also applicable, in particular measurement of splitting in the ODMR spectrum. In this case the sensitivity of the fNDs embedded in ENFs for magnetometry can be estimated using the equation. [27]:

$$\eta = \frac{4h\delta}{3\sqrt{3}g_{nv}\mu_B R\sqrt{N}} \qquad (1)$$

Where η represents sensitivity, δ is the full width at half maximum of the ODMR peak, R represents the experimental ODMR contrast, $g_{nv}$ is the g-factor of the NV centre, $\mu_B$ is the Bohr magneton and N is the number of collected photons per second. Using Equation 1 and measurements of PL from a single ND within the FOV represented in figure 2D, the sensitivity of the fND embedded ENFs is estimated to be 3.4μT/√Hz for a 2s integration time. Higher sensitivity could be achieved through an increase in the number of photons collected via improvement in light collection efficiency, in the number of embedded fNDs and use of a less optically scattering polymer. Enhanced sensitivity is also possible through reduction of the ODMR linewidth, δ, by improvement in the coherence properties of the NVs, potentially achievable using higher purity $^{12}$C enriched diamond[28] since the inhomogeneous dephasing of the NV electron spin is influenced primarily by the composition of the surrounding spin bath in the diamond lattice[4]. Sensitivity could also be improved by enhancing the contrast, R, which theoretically is significantly higher (~ 30 %) than the one measured here (2 % to 3 %) however, in practise experimental contrast rarely exceeds 10-15% due to broadening from the strain environment of the ND crystal lattice. More efficient delivery of MWs would lead to higher contrast, indeed the literature reports a 5 fold improvement in contrast using a micro strip antenna compared with a copper wire as used in this report[29].

Improvements in the detection sensitivity of the presented measurement platform could pave the way for the implementation of spatiotemporal electrophysiological measurements on live cells in a physiological-like environment. Indeed, the concept of electrophysiological measurements in cells using NV centers has been explored theoretically and experimentally [30,31]. Whilst still in its infancy compared to existing methods including patch clamping, whereby deformation of the cell membrane is achieved using a hybrid electrode/pipette tip[32] and fluorescence methods based on the use of genetically engineered voltage sensitive proteins [33], or electrochromic and FRET based dyes [34], electrical sensing using NV centers has made significant advances in recent years. Moreover, NV based approaches overcome many of the drawbacks of existing methods such as the inherent invasiveness, low throughput and lack of spatial information associated with patch clamping as well as difficulties associated with fluorescence based sensors, namely the requirement for genetic manipulation, or observed cytotoxicity of many fluorescence based probes[35]. To date, a number of studies have considered in detail the different experimental parameters necessary for maximising measurement sensitivity in NV based sensing of neuronal action potentials[30,31,36]. For example, a recent demonstration of optomagnetic detection of single axonal potentials employed a MW lock in detection method to manipulate the NV spin state, citing sensitivity at a distance of 10 µm from the sample and by integrating the signal over a relatively large surface area (~8mm$^2$) of 15pT/√Hz. This was more than sufficient to observed the peak fields during axonal firing measured in the region of a few nT[31]. However, the aforementioned study utilises an experimental setup consisting of dissected marine worms and squid, in which a central axon running the length of the organism is stimulated using external electrodes and measured in proximity to a grown NV diamond layer. As is stressed in this report, a key step forward for NV sensing in biological systems will be the ability to measure endogenous electrical activity in cells using an experimental setup that recapitulates the physiological cell nice.

*Growth, differentiation and acquisition of ODMR spectra on NSCs cultured on fND embedded ENFs*

To examine the potential for fND embedded nanofibers to be used in biological systems an investigation of whether NSCs could be grown and differentiated into their daughter cells on the fibers was carried out. NSCs grown on the ENFs for 7 days displayed a morphology typical of the expected differentiated cell phenotypes. The viability of the differentiated NSCs on the fND embedded ENFs was determined from the live dead assay to be 85% which is in the range previously reported for differentiated NSCs cultured on glass cover slips (Figures 6a, 6b, 6 and 6j). The percentage of apoptotic cells, as assessed by the numbers of pyknotic nuclei, was low with percentages of 4 % for cells cultured on the ENFs and 2.5 % for cells cultured on the glass coverslip observed (Figure 6i). Combined these results demonstrate cell viability was not compromised following culture on fND embedded ENFs.

Figure 6c-h shows fluorescent images of immunostained cells cultured on glass coverslips and fND embedded ENFs. Marker expression confirms cells underwent successful differentiation into their daughter cell types (astrocytes, neurons and oligodendrocytes) when cultured on glass and ENFs (Figures 7a to 7f). Figure 6k indicates similar levels of differentiation into astrocytes occurred for both culture substrates. However, marker expression for neurons and oligodendrocytes slight higher on ENFs as compared to glass (Figures 6 l and m). In addition, oligodendrocytes appeared to display more extensive ramifications and broader surface area when grown on the ENFs as compared to glass. The observed differences in marker expression demonstrate the usefulness of ENFs to promote NSC differentiation into functional neurons and oligodendrocytes as seen *in vivo*. Overall, the fND embedded ENFs provided a suitable substrate for cell growth and differentiation, with no adverse effect on cell function observed.

To demonstrate the utility of the fND embedded ENFs for quantum sensing in live cell environments, cultures of neural stem cells were grown for a period of 21 days, at which point mature networks were assumed to have formed. We confirmed the presence of these active networks via live cell staining with fluo5AM, a fluorescence based probe that exhibits higher fluorescence upon binding $Ca^{2+}$. $Ca^{2+}$ imaging is widely used as a surrogate marker for neuronal electrical activity, as the signalling mechanism itself is mediated by cellular action potentials [37]. Figure 7a shows time lapse imaging of calcium flux observed in neuronal cultures after 21 days culture. Dual fluorescence imaging (top left panel) shows a snapshot of calcium stained neurons growing on the fND embedded fibres, further demonstrating the suitability of these substrates for maintaining viable populations of cells for extended *in vitro* culture periods. The coloured heat map images in figure 7a track the progression of calcium flux through the cellular network over a period of 8 seconds (supplementary video 1). Inset white arrows highlight regions where the flux of calcium is most prominent for each timestamp. Figure 7b shows a population of single neurons, which were qualitatively identified via their characteristic cell shape and dendritic outgrowths. The activity was confirmed using time-lapse imaging of intracellular calcium flux (supplementary video 2), after which ODMR spectra where then obtained for single fNDs in close proximity to identified neurons in two separate regions (figure 7bi, inset blue and yellow boxes). Figure 7biii&iv show regions around which fNDs were selected in proximity to the neuron cell body and dendrites (inset yellow boxes). fNDs in 7biii are pseudo coloured grey to allow ease of viewing of the fNDs that were analysed. The two graphs in figure 7 show ODMR spectra for the aforementioned regions. The acquisition of ODMR spectra from cellular aggregates of fND ensembles is demonstrated in figure 7c. The aggregate was identified, the activity assessed (supplementary video 3) and the ODMR spectrum acquired as previously described. The resultant spectra from fNDs residing underneath the cell aggregate has a significantly higher signal to noise than spectra in figure 7b, since single to noise is proportional to $\sqrt{n}$, where n is the number of measurements made, and the ensemble contains a far greater number of fNDs. The results presented in figure 7c represent potentially a more promising approach to detect endogenous electrical activity from neurons compared with a single cell/single fND approach, since the ODMR spectra from the fND ensembles compared with single fNDs provide a far higher signal to noise, but also that electrical activity in aggregates of neuronal cells will likely provide a stronger signal compared with single cells.

It is noted that whilst detection of endogenous electrical signals from cells is not presented in this report, emphasis must be placed on the technical challenges facing such measurements. Furthermore, the validation of active neural circuits in cells grown on fND embedded ENFs using quantum sensing protocols would represent a significant advancement developing methods in electrophysiology. As previously discussed, further optimisation of the fabrication procedure and improvements to the experimental setup and antenna design could significantly improve the sensitivity of this sensing platform, and thus realise the true utility of this bio-sensing approach. In addition, modifications to the acquisition protocol that allow simultaneous imaging of calcium flux and ODMR acquisition, will not only provide concrete validation of this novel electrophysiology approach, but could elucidate important information about the onset of calcium flux in neurons following the initiation of cellular action potentials. Whilst this latter concept has previously been demonstrated using calcium dyes on multi electrode arrays[38], the implementation of quantum sensing for this application would provide a first demonstration of fully quantitative spatiotemporal electrophysiology in single cell and multicellular environments.

*Conclusions*

A quantum sensing platform that mimics the nanoscale architecture and topography of the cell niche has been presented. This work identified and demonstrated formulations and fabrication protocols for the production of polymer ENFs with fNDs uniformly embedded along the fibers. The embedded fNDs were used to successfully acquire ODMR spectra, perform longitudinal spin relaxometry and detect external magnetic fields. The ENFs supported the growth and differentiation of NSCs, with viability after 7 days of culture comparable to cells cultured on control glass coverslips. Further the ENF substrates supported the formation of active neural networks, and could be used for quantum sensing in single and multicellular environments. Overall, this work has provided a new approach to NV based quantum sensing in active neural networks and in doing so addressed the need for high precision biosensors integrated in environments that mimic *in vivo* cellular niches. Furthermore, it is envisaged that this platform could serve a number of applications in studying *in vitro* biological environments, whereby simultaneously monitoring of electrical activity, free radical production and local changes in environmental temperature is attainable. Future work seeks the realisation of NV based detection of endogenous electromagnetic fields emanating from live cells. To this end, work is currently underway to improve the sensitivity of the fND embedded ENFs and deploy them for detection of endogenous electrical signals from electrically active neurons and isolated chick cardiomyocytes. Finally, the demonstration of quantum sensing in electrospun scaffolds that mimic the nanoscale properties of the cell niche will be of particular interest to those working in stem cell biology and tissue engineering.

*Acknowledgements*

The authors wish to acknowledge the European Research Council (ERC) for funding this work through the ERC Consolidator Award, *TransPhorm*, grant number 23432094.

*References*

1. Acosta, V. & Hemmer, P. Nitrogen-vacancy centers: Physics and applications. *MRS Bull.* **38,** 127–130 (2013).

2. Casola, F., Van Der Sar, T. & Yacoby, A. Probing condensed matter physics with magnetometry based on nitrogen-vacancy centres in diamond. *Nature Reviews Materials* **3,** (2018).

3. Schirhagl, R., Chang, K., Loretz, M. & Degen, C. L. Nitrogen-Vacancy Centers in Diamond: Nanoscale Sensors for Physics and Biology. *Annu. Rev. Phys. Chem* **65,** 83–105 (2014).

4. Rondin, L. *et al.* Magnetometry with nitrogen-vacancy defects in diamond. *Rep Prog Phys* **77,** 56503 (2014).

5. Le Sage, D. *et al.* Optical magnetic imaging of living cells. *Nature* **496,** 486–489 (2013).

6. Steinert, S. *et al.* Magnetic spin imaging under ambient conditions with sub-cellular resolution. *Nat. Commun.* **4,** (2013).

7. Barry, F. *et al.* Correction for Barry et al., Optical magnetic detection of single-neuron action potentials using quantum defects in diamond. *Proc. Natl. Acad. Sci.* **114,** E6730–E6730 (2017).

8. Kucsko, G. *et al.* Nanometre-scale thermometry in a living cell. *Nature* **500,** 54–58 (2013).

9. Simpson, D. A. *et al.* Non-Neurotoxic Nanodiamond Probes for Intraneuronal Temperature Mapping. *ACS Nano* **11,** 12077–12086 (2017).

10. Zhu, Y. *et al.* The biocompatibility of nanodiamonds and their application in drug delivery systems. *Theranostics* (2012). doi:10.7150/thno.3627

11. Geissler, M. *et al.* Primary Hippocampal Neurons, Which Lack Four Crucial Extracellular Matrix Molecules, Display Abnormalities of Synaptic Structure and Function and Severe Deficits in Perineuronal Net Formation. *J. Neurosci.* (2013). doi:10.1523/JNEUROSCI.3275-12.2013

12. Wilson, C. L., Hayward, S. L. & Kidambi, S. Astrogliosis in a dish: Substrate stiffness induces astrogliosis in primary rat astrocytes. *RSC Adv.* **6,** 34447–34457 (2016).

13. Lantoine, J. *et al.* Matrix stiffness modulates formation and activity of neuronal networks of controlled architectures. *Biomaterials* **89,** 14–24 (2016).

14. Boothe, S. D. *et al.* The Effect of Substrate Stiffness on Cardiomyocyte Action Potentials. *Cell Biochem. Biophys.* **74,** 527–535 (2016).

15. Lim, S. H. & Mao, H. Q. Electrospun scaffolds for stem cell engineering. *Advanced Drug Delivery Reviews* (2009). doi:10.1016/j.addr.2009.07.011

16. Li, J. *et al.* Human Pluripotent Stem Cell-Derived Cardiac Tissue-like Constructs for Repairing the Infarcted Myocardium. *Stem Cell Reports* **9,** 1546–1559 (2017).

17. Bourke, J. L., Coleman, H. A., Pham, V., Forsythe, J. S. & Parkington, H. C. Neuronal Electrophysiological Function and Control of Neurite Outgrowth on Electrospun Polymer Nanofibers Are Cell Type Dependent. *Tissue Eng. Part A* **20,** 1089–1095 (2014).

18. Weightman, A., Jenkins, S., Pickard, M., Chari, D. & Yang, Y. Alignment of multiple glial cell populations in 3D nanofiber scaffolds: toward the development of multicellular implantable scaffolds for repair of neural injury. *Nanomedicine* **10,** 291–5 (2014).

19. Lai, H. J. *et al.* Tailored design of electrospun composite nanofibers with staged release of multiple angiogenic growth factors for chronic wound healing. *Acta Biomater.* (2014). doi:10.1016/j.actbio.2014.05.001

20. Xie, Z. *et al.* Dual growth factor releasing multi-functional nanofibers for wound healing. *Acta Biomater.* (2013). doi:10.1016/j.actbio.2013.07.030

21. Xu, Q., Jin, L., Li, C., Kuddannayai, S. & Zhang, Y. The effect of electrical stimulation on cortical cells in 3D nanofibrous scaffolds. *RSC Adv.* **8,** 11027–11035 (2018).

22. Sokolov, P. A., Belousov, M. V., Bondarev, S. A., Zhouravleva, G. A. & Kasyanenko, N. A. FibrilJ: ImageJ plugin for fibrils' diameter and persistence length determination. *Comput. Phys. Commun.* (2017). doi:10.1016/j.cpc.2017.01.011

23. Simpson, D. A. *et al.* Non-Neurotoxic Nanodiamond Probes for Intraneuronal Temperature Mapping. *ACS Nano* **11,** 12077–12086 (2017).

24. Ermakova, A. *et al.* Detection of a few metallo-protein molecules using color centers in nanodiamonds. *Nano Lett.* **13,** 3305–3309 (2013).

25. Tetienne, J. P. *et al.* Spin relaxometry of single nitrogen-vacancy defects in diamond nanocrystals for magnetic noise sensing. *Phys. Rev. B - Condens. Matter Mater. Phys.* **87,** (2013).

26. Rendler, T. *et al.* Optical imaging of localized chemical events using programmable diamond quantum nanosensors. *Nat. Commun.* **8,** (2017).

27. Dréau, A. *et al.* Avoiding power broadening in optically detected magnetic resonance of single NV defects for enhanced dc magnetic field sensitivity. *Phys. Rev. B - Condens. Matter Mater. Phys.* (2011). doi:10.1103/PhysRevB.84.195204

28. Balasubramanian, G. *et al.* Ultralong spin coherence time in isotopically engineered diamond. *Nat. Mater.* (2009). doi:10.1038/nmat2420

29. Qin, L. *et al.* Near-field microwave radiation function on spin assembly of nitrogen vacancy centers in diamond with copper wire and ring microstrip antennas. *Jpn. J. Appl. Phys.* (2018). doi:10.7567/JJAP.57.072201

30. Hall, L. T. *et al.* High spatial and temporal resolution wide-field imaging of neuron activity using quantum NV-diamond. *Sci. Rep.* **2,** 401 (2012).

31. Barry, F. *et al.* Optical magnetic detection of single-neuron action potentials using quantum defects in diamond. *Proc. Natl. Acad. Sci.* **114,** E6730–E6730 (2017).

32. Sakmann, B. & Neher, E. Patch Clamp Techniques for Studying Ionic Channels in Excitable Membranes. *Annu. Rev. Physiol.* (1984). doi:10.1146/annurev.ph.46.030184.002323

33. Liu, P., Grenier, V., Hong, W., Muller, V. R. & Miller, E. W. Fluorogenic Targeting of Voltage-Sensitive Dyes to Neurons. *J. Am. Chem. Soc.* **139,** 17334–17340 (2017).

34. Miller, E. W. *et al.* Optically monitoring voltage in neurons by photo- induced electron transfer through molecular wires. *Proc. Natl. Acad. Sci. U. S. A.* **109,** 2114–2119 (2011).

35. Rowland, C. E., Brown, C. W., Medintz, I. L. & Delehanty, J. B. Intracellular FRET-based probes: A review. *Methods and Applications in Fluorescence* (2015). doi:10.1088/2050-6120/3/4/042006

36. Karadas, M. *et al.* Feasibility and resolution limits of opto-magnetic imaging of neural network activity in brain slices using color centers in diamond. *Sci. Rep.* **8,** (2018).

37. Grienberger, C. & Konnerth, A. Imaging Calcium in Neurons. *Neuron* (2012). doi:10.1016/j.neuron.2012.02.011

38. Shew, W. L., Bellay, T. & Plenz, D. Simultaneous multi-electrode array recording and two-photon calcium imaging of neural activity. *J. Neurosci. Methods* (2010). doi:10.1016/j.jneumeth.2010.07.023

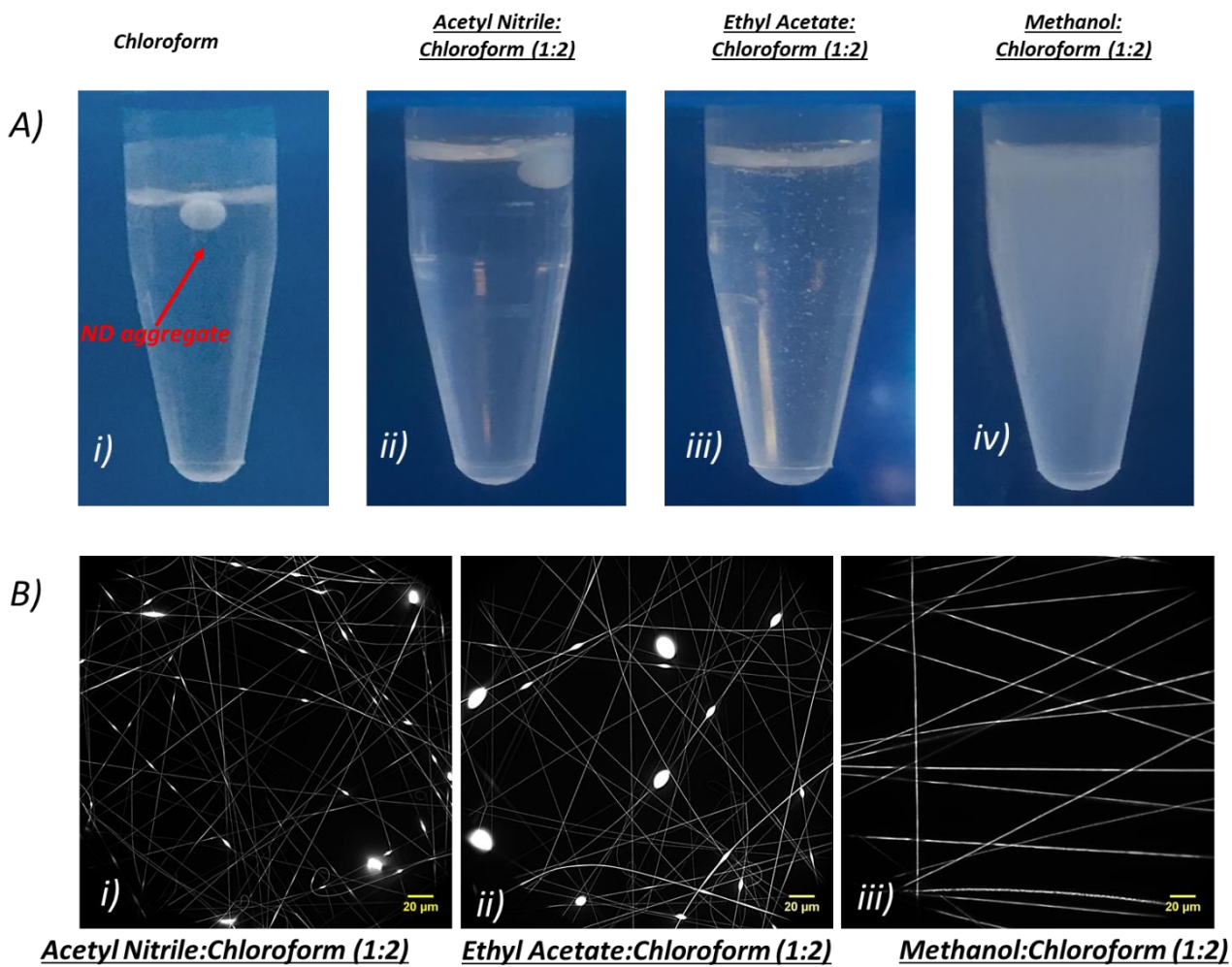

**Fig1:** Dependence of fND dispersion and ENF formation on solvent composition. **a)** Suspensions of fNDs in (i) chloroform, (ii) acetyl nitrile:chloroform (1:2), (iii) Ethyl Acetate:Chloroform (1:2), (iv) Methanol:Chloroform (1:2); **b)** Fluorescence microscopy images of Rhodamine-B labelled ENFs spun with different solvent compositions (i) acetyl nitrile:chloroform (1:2), (ii) Ethyl Acetate:Chloroform (1:2), (iii) Methanol:Chloroform (1:2).

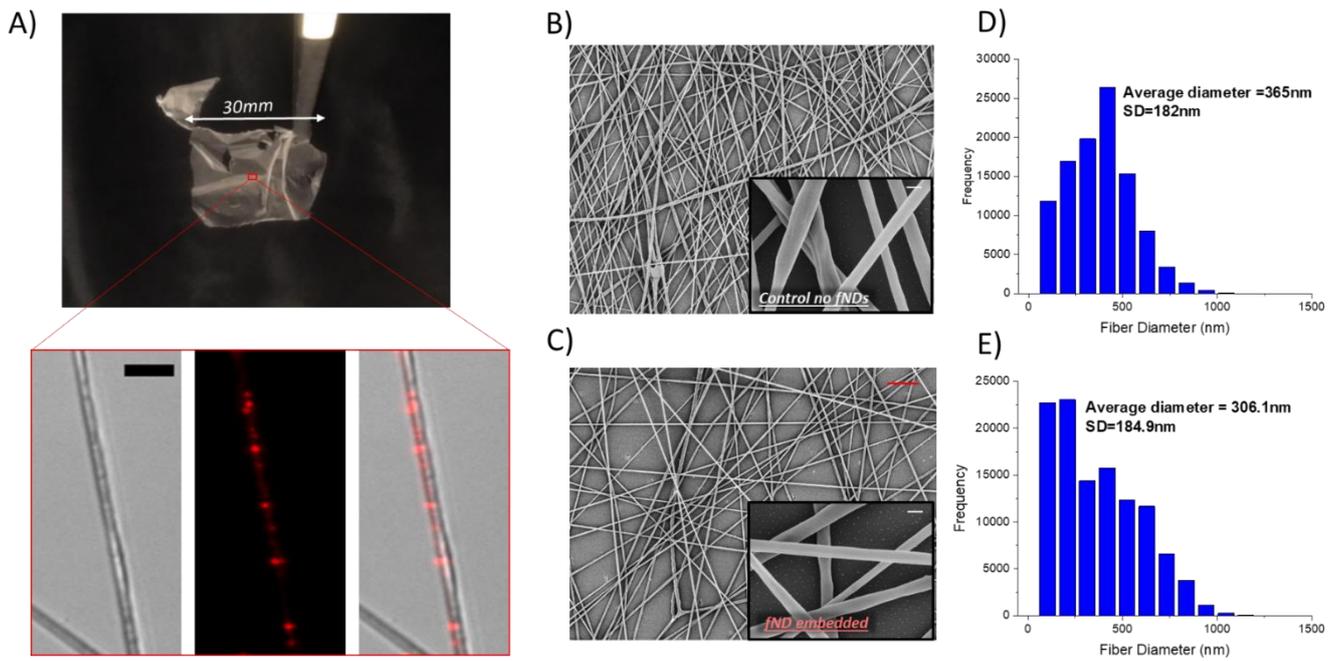

**Fig2: Morphological characterisation of fND embedded ENFs**. **a)** Macroscopic image of whole fND embedded ENF matt with inset showing fNDs and ENFs visualised with widefield DIC and fluorescence microscopy, **b & c**) Scanning electron microscopy images of control (**b**) and fND embedded fibers (**c**), **d & e**) Histograms displaying the distribution of fiber diameters for control and fND embedded fibres.

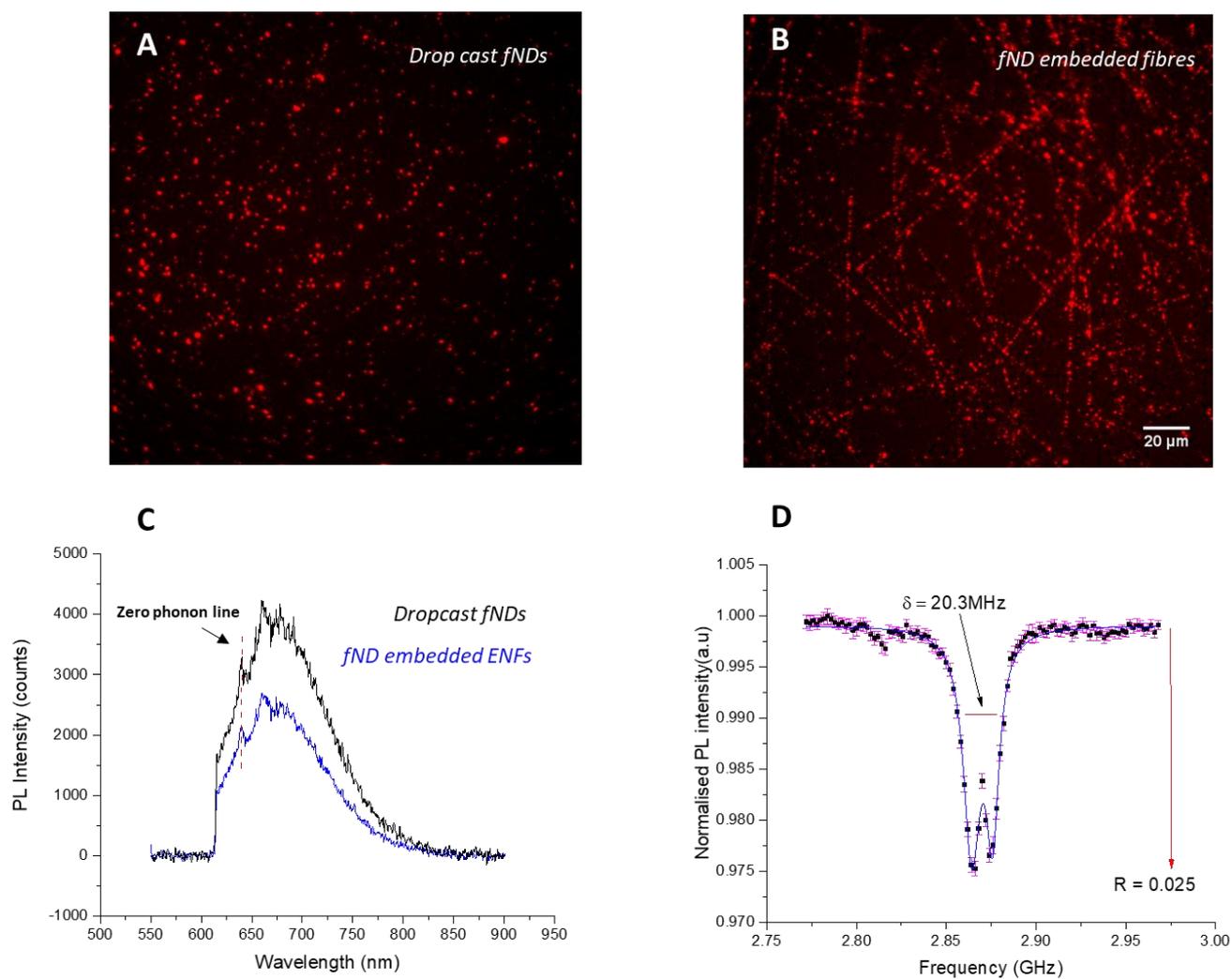

**Fig3: Comparison of emission spectra and ODMR signals between drop cast fNDs and fND embedded ENFs**. Widefield fluorescence imaging of drop cast fNDs **(a)** and fND embedded fibers **(b)**, emission spectra for dropcast and fND embedded ENFs **(c)** and ODMR spectrum acquired from fND embedded ENFs.

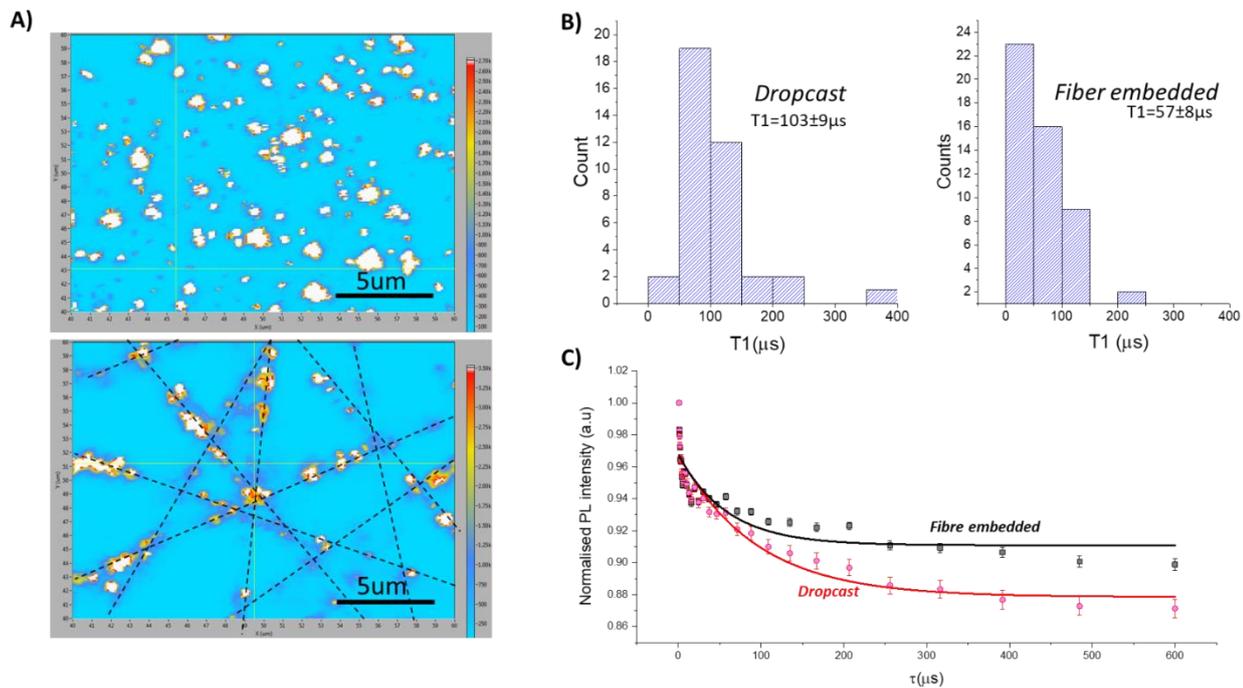

**Fig4: Comparison of longitudinal (T1) relaxation times between drop cast and fND embedded ENFs**. (**a**) Intensity maps of T1 relaxation times from dropcast (top panel) and fND embedded ENFs (bottom panel), (**b**). Histograms of relaxation times from individual regions within the intensity maps displayed in (a), Time dependent change in PL intensity for dropcast (**c**) and fibre embedded (**d**) fNDs.

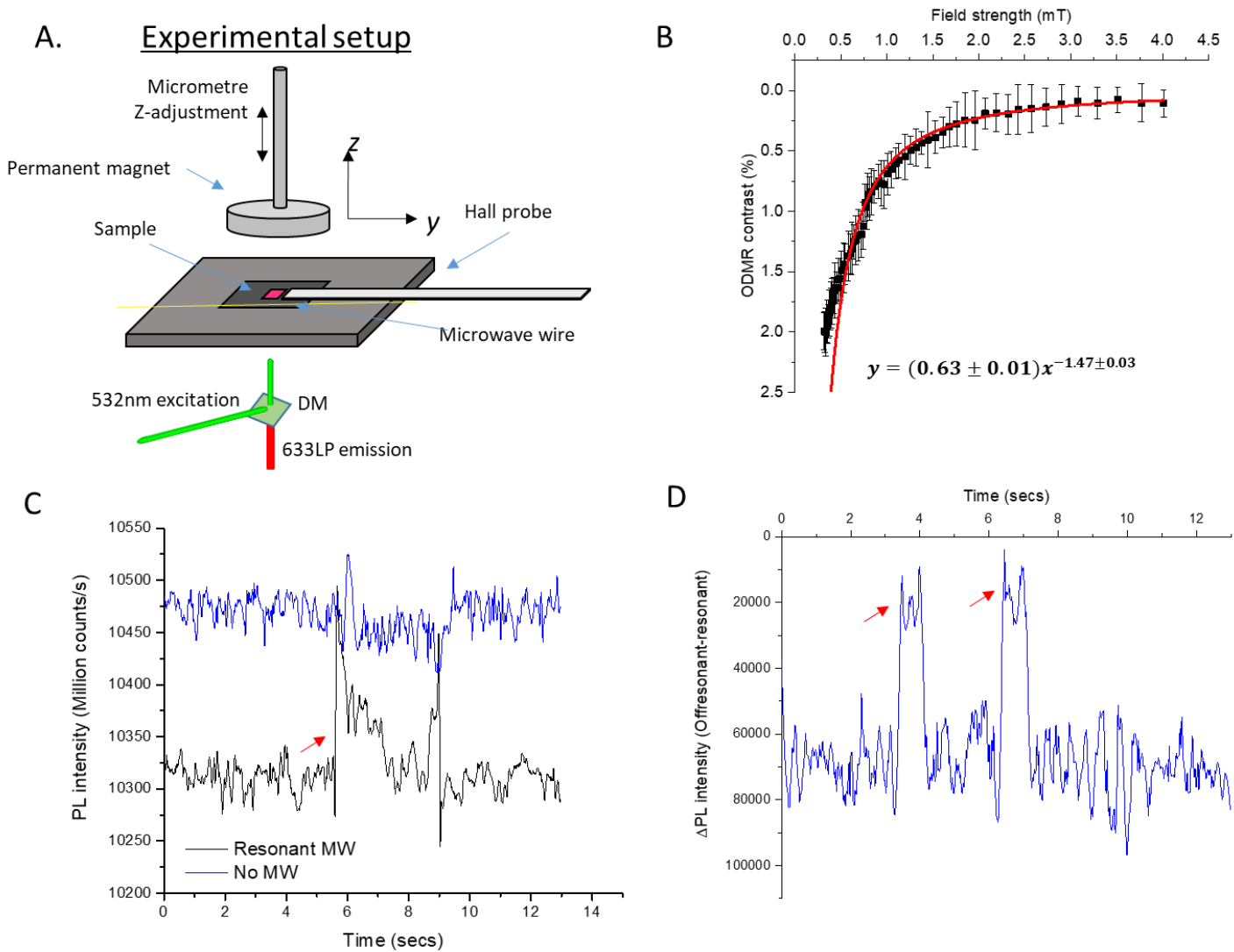

**Fig5: Detection of external magnetic fields. (a)** Schematic of experimental set up for obtaining magnetic field strength calibration curve and ODMR contrast, **(b)** Calibration curve relating magnetic field strength to ODMR contrast in fND embedded fibres. **(c & d)** Time resolved magnetic field detection using fND embedded fibres. Inset red arrow shows the point at which a field is applied by momentarily holding a permanent magnet in close proximity to the fibres.

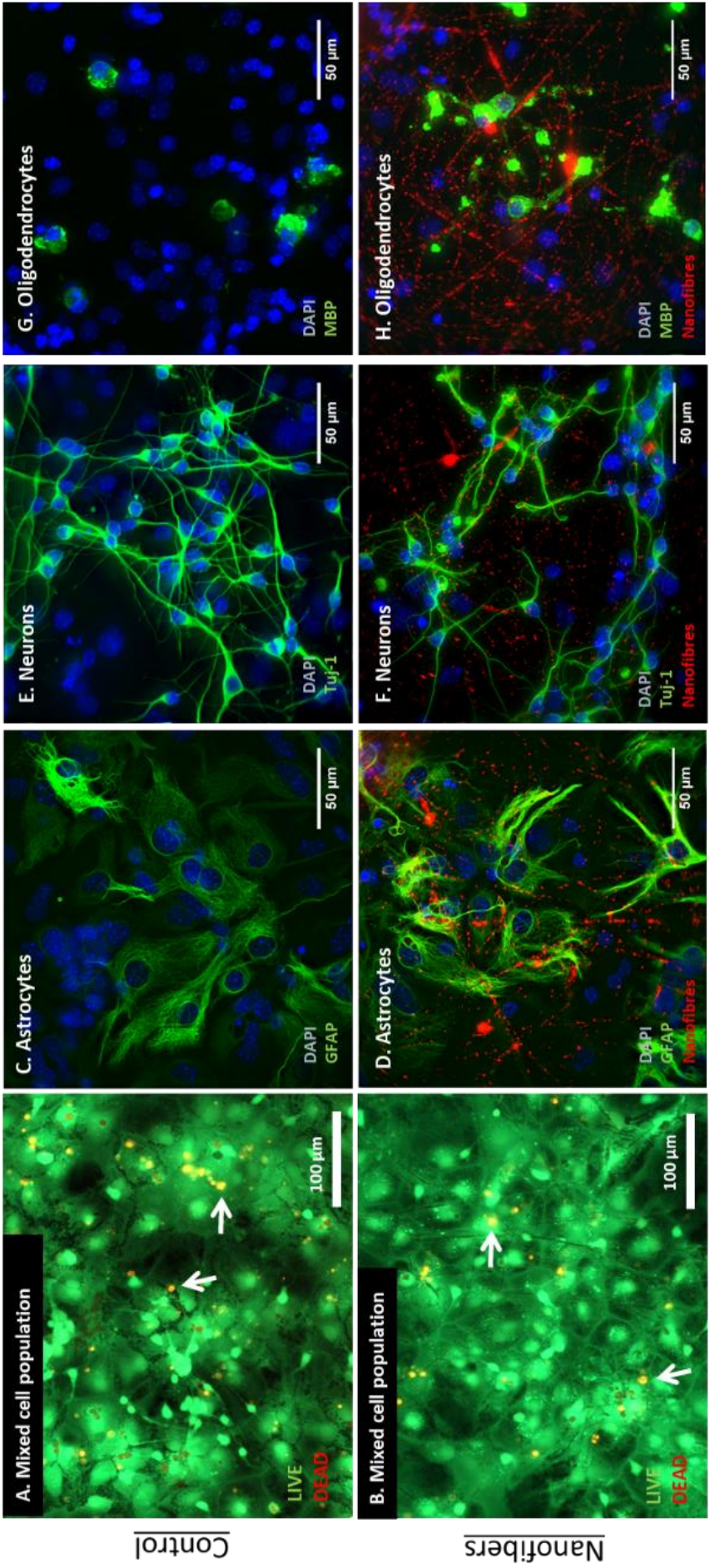
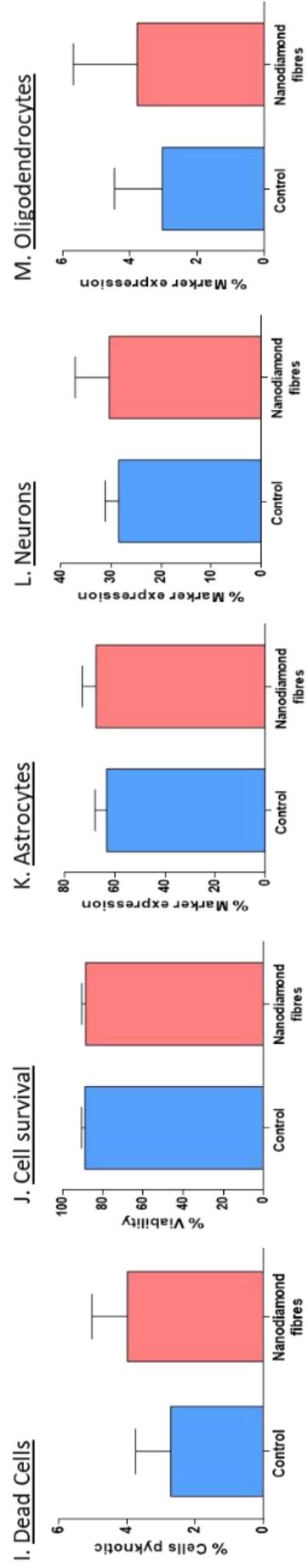

**Fig6: Characterisation of neural cell phenotypes in differentiating NSC after 7 days' culture on ND embedded nanofibers.** Fluorescence microscopy images of live/dead staining in neuronal cells cultured on glass **(a)** and nanofibers **(b)** for 7 days, green indicates live cells and red indicates pyknotic nuclei in apoptotic cells. **(c to h)** Fluorescent microscopy images of immunostained cells grown on glass coverslips (c, e & g ) and fNDs embedded ENFs (d, f & h). In all images cell nuclei (blue) can be identified from DAPI staining. Green indicates astrocyte marker GFAP expression in (c) and (d), neuronal marker Tuj-1 expression in (e) and (f) and oligodendrocyte marker MBP expression in (g) and (h). fNDs are seen in (d) to (h) in the red channel. Quantification of cell viability **(J)** and percentage of Pyknotic nuclei **(i)** for cells cultured on glass coverslips and fND embedded ENFs, error bars represent standard error of the mean across multiply experimental repeats (n=3). Quantification of marker expression for cells grown on glass coverslip controls and fND embedded ENFs (**k to m**), error bars represent standard error of the mean across multiply experimental repeats (n-=3).

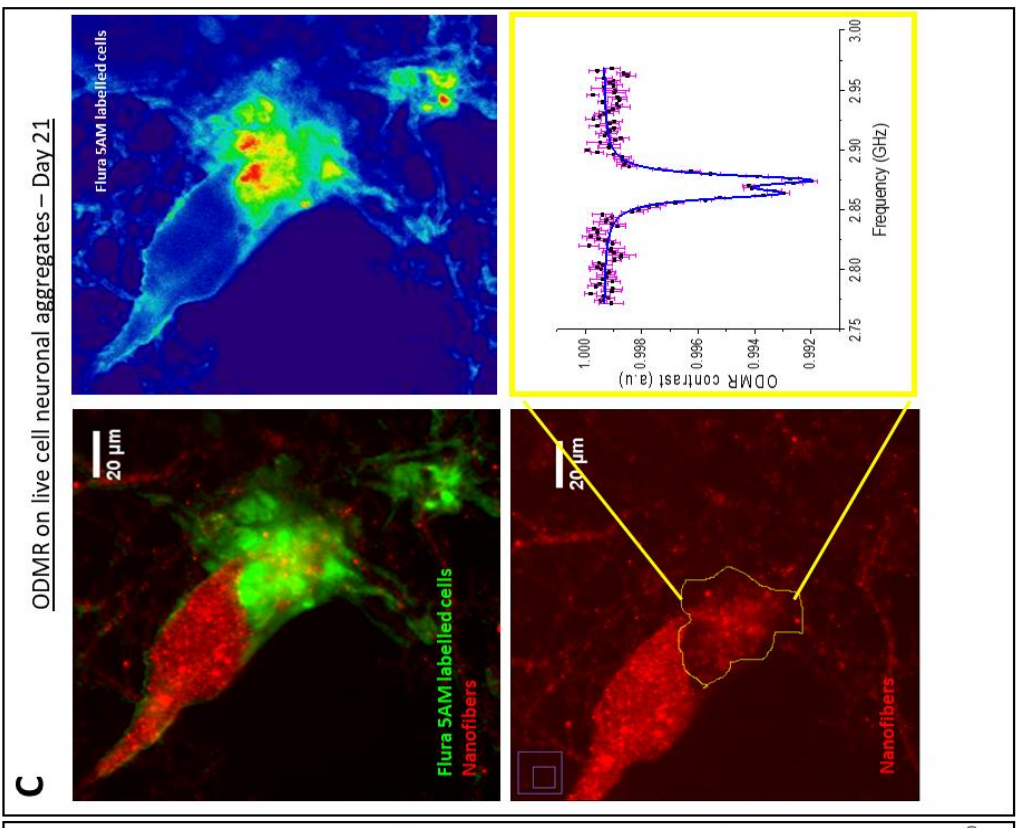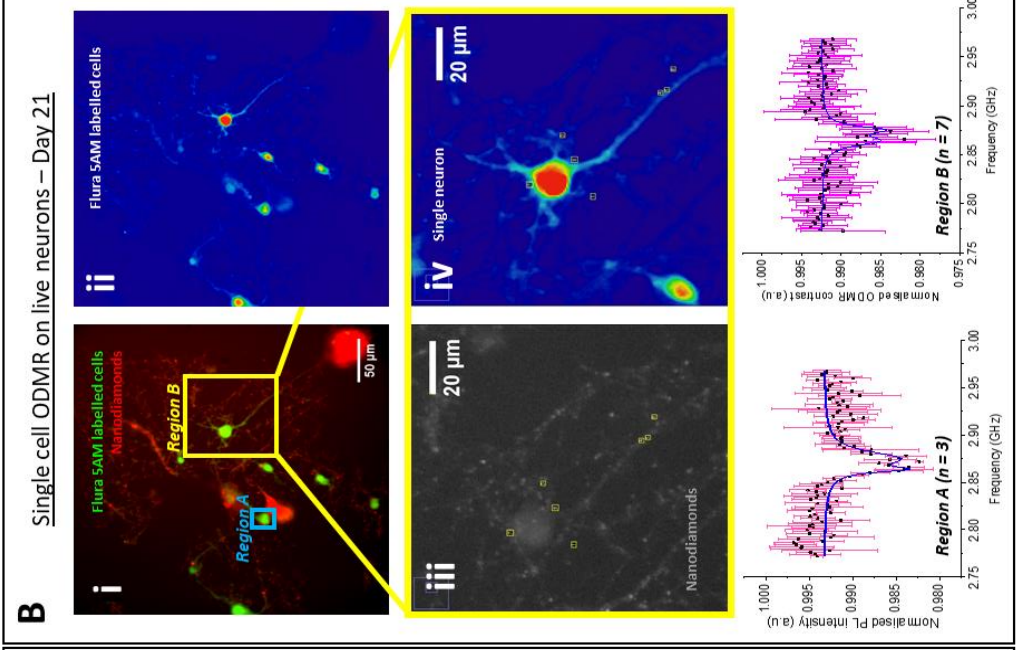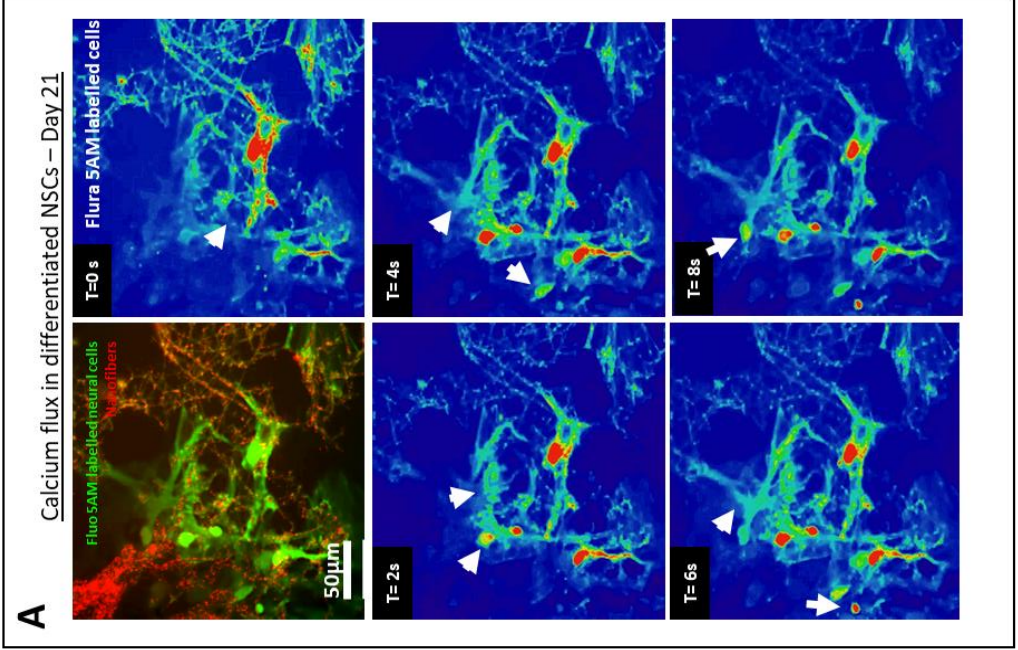

**Fig6: Real time acquisition of neuronal activity in day 21 differentiated NSCs grown on fND embedded ENFs.** Time lapse data for calcium fluxes in neural networks in day 21 cultured NSCs (**a**). Dual fluorescence images show fNDs embedded in ENFs (red) and a mixed population of neural cells stained with a fluo-5AM to visualise calcium flux (green). Heat map time lapse imaging shows the flux of calcium over a period of 8 seconds. Inset white arrows highlight areas where flux is most prominent. Single neurons were identified qualitatively (**b**i&ii) via characteristic cell shape and dendrite growth. Snapshots of time lapse data of calcium flux was used to identify active cells in two separate regions (inset blue and yellow boxes). Higher magnification images (expanded yellow box) show fNDs located close to the cell body and dendrites (**b**iii&iv), on which ODMR of single fNDs was subsequently performed (bottom graphs). Larger cell aggregates (**c**) and ODMR of fND ensembles was performed in bulk in a region co-localised with areas with a high calcium flux (traced yellow region). Error bars for ODMR spectra in **b** and **c** represent standard deviation from the mean of 20ms exposures averaged over a 2s integration time.